\documentclass[journal,twoside,twocolumn,letterpaper]{IEEEtran}

\usepackage{color}
\usepackage[latin1]{inputenc}
\usepackage{cite}
\usepackage{mathtools}
\usepackage{amssymb}
\usepackage{algorithm}
\usepackage{algpseudocode}
\usepackage[artemisia]{textgreek}
\usepackage{url,cite}

\usepackage{tikz}
\usetikzlibrary{decorations.pathmorphing} 
\usetikzlibrary{fit}					
\usetikzlibrary{backgrounds}	
\usetikzlibrary{shapes}
\usetikzlibrary{positioning}
\usetikzlibrary{shadows}
\usetikzlibrary{decorations.pathmorphing}
\usetikzlibrary{decorations.shapes}
\usepackage{multirow}
\usepackage{dcolumn}
\usepackage{pgfplots}
\pgfplotsset{compat=1.3}

\newtheorem{theorem}{Theorem}
\newtheorem{lemma}{Lemma}

\newtheorem{corollary}{Corollary}

\begin{document}
\allowdisplaybreaks

\pagestyle{headings}
\title{On Interference Dynamics in Mat\'ern Networks}
\author{Udo Schilcher, Jorge F. Schmidt, and Christian Bettstetter~\IEEEmembership{Senior~Member,~IEEE}
	\thanks{
			Udo Schilcher is with Lakeside Labs GmbH, Klagenfurt, Austria, E-Mail: \texttt{schilcher@lakeside-labs.com}.
			Jorge F. Schmidt, and Christian Bettstetter are with the Institute of Networked and Embedded Systems, University of Klagenfurt, Austria, E-mail: \texttt{\{jorge.schmidt,christian.bettstetter\}@aau.at}.
	}
}

\markboth{}{}
\thispagestyle{empty}

\makeatletter
\newcommand{\nvast}{\bBigg@{3}}
\newcommand{\vast}{\bBigg@{3.5}}

\makeatother

\newcommand{\etal}{{et al.\;}}

\newcommand{\expect}[1]{{{\mathbb{E}}\!\left[{#1}\right]}}
\newcommand{\var}[1]{{{\rm var}\!\left[{#1}\right]}}
\newcommand{\cov}[2]{{{\rm cov}\!\left[{#1},{#2}\right]}}
\newcommand{\cor}[2]{{\rho\!\left[{#1},{#2}\right]}}

\newcommand{\eulergamma}{\gamma_{\mathrm{eul}}}
\newcommand{\Ei}[1]{E_i\left(#1\right)}

\newcommand{\R}{\mathbb{R}}
\newcommand{\dd}{{\rm d}}
\newcommand{\uniform}{\mathbb{U}(0,1)}

\newcommand{\ppp}{\Phi_{\rm p}}
\newcommand{\tpp}{\Phi}
\newcommand{\densppp}{\lambda_{\rm p}}
\newcommand{\mc}{\range^2\pi}
\newcommand{\interf}{I}
\newcommand{\fad}[1]{h_{#1}^2}
\newcommand{\fadsq}[1]{h_{#1}^4}
\newcommand{\ploss}[1]{\ell\left(\left\|#1\right\|\right)}
\newcommand{\plosssc}[1]{\ell\left(#1\right)}
\newcommand{\plosssq}[1]{\ell^2(\|#1\|)}
\newcommand{\channel}[1]{\fad{#1}\,\ploss{#1}}
\newcommand{\indic}[2]{\gamma_{#1}(#2)}
\newcommand{\prdensii}[1]{\rho^{(2)}(#1)}
\newcommand{\dens}{\lambda}

\newcommand{\range}{d}
\newcommand{\Gh}{\Gamma_\range(r)}
\newcommand{\gh}{\gamma_\range(r)}
\newcommand{\ghpot}[1]{\gamma^{#1}_\range(r)}

\newcommand{\pa}{p_1}
\newcommand{\pab}{p_{12}}
\newcommand{\papa}{p_{1/1}}
\newcommand{\papb}{p_{1/2}}

\newcommand{\areac}{A_c}
\newcommand{\areax}{A_x}
\newcommand{\areay}{A_y}

\newcommand{\cir}{\mathcal{C}}

\sloppy

\maketitle
\thispagestyle{empty}
\begin{abstract}
A thorough understanding of the temporal dynamics of interference in wireless networks is crucial for the design of communication protocols, scheduling, and interference management. This paper applies stochastic geometry to investigate interference dynamics for the first time in a network with nodes that use carrier sense multiple access. This type of networks is approximated by a Mat\'ern hard-core point process of type~II with Nakagami fading. We derive and analyze expressions for the variance, covariance, and correlation of the interference power at a given point in space. 
Results show that even though the commonly used Poisson approximation to carrier sense multiple access may have the same average interference than the Mat\'ern model, the three interference dynamics measurements investigated behave significantly different.
\end{abstract}

\section{Introduction}\label{sec:intro}

\subsection{Motivation}
The modeling and analysis of interference in wireless networks by means of stochastic geometry \cite{stoyan95} has become popular in the course of the past 15~years (see \cite{net:Haenggi09jsac,baccelli09:vol2,haenggi13:book}). The majority of work in this domain uses simple modeling assumptions for reasons of mathematical tractability. A common network model includes uniformly randomly distributed nodes employing slotted ALOHA for medium access and Poisson arrival of transmission demands. This leads to uniformly distributed senders for interference analysis, which can be modeled by a Poisson point process (PPP). Such networks are sometimes called {\it Poisson networks\/}. They facilitate the derivation of mathematical expressions while retaining important network properties; they are well understood in the communications theory community with several results available in the literature (see~\cite{haenggi13:book,baccelli09:vol2,schilcher15:tit,haenggi13:div-poly,schilcher2013coop,ganti09:interf-correl,tanbourgi14:mrc}).

In practice, however, many computer and communication networks do not access the shared medium in an ALOHA style but perform some type of {\it carrier sensing\/}. A classical example is Carrier Sense Multiple Access (CSMA), in which each node senses the medium and only starts to send a message if the medium is idle (otherwise backs off according to some rule and tries to send at a later instant) with the goal to reduce the number of message collisions. CSMA is also the basis for medium access control in IEEE 802.11~\cite{7786995}. Although such medium sensing is not captured by Poisson networks, CSMA networks are approximated in most analytical studies by Poisson networks \cite{Kaufman2013,Ye2013,ElSawy2014,Lin2014b,Schmidt2015}. This approximation has been shown to be accurate for first-order statistics of the interference power, i.e., statistics on interference at one point in space, such as expected value and variance (see \cite{Ye2013,ElSawy2014,nguyen07:csma}). Its suitability for higher-order statistics of the interference (i.e., at several points in time or space) is, however, still unknown.

A better, very natural choice to model senders with carrier sensing is a Mat\'ern hard-core point process (MPP) \cite{matern86}. It introduces a guard circle around each sender, in which no node is allowed to send~\cite{haenggi11:mean-interf,baccelli09:vol2}. This resembles the medium sensing and is suited for any kind of wireless network with spatial reservation of a transmission floor. Such {\it Mat\'ern networks} are still only approximations for CSMA networks because real CSMA protocol implementations are more complex. For example, there is a chance that two close nodes send simultaneously due to imperfect sensing or hidden node effects. MPP assumes perfect sensing, thus leading to a slight underestimation of interference. Nevertheless, the approximation by MPP is much closer to CSMA than any PPP model and brings the theory of interference calculus a step forward in the direction of more realistic models. From a more general perspective, only few analytical results are available for MPPs, severely limiting the accurate analysis of many modern wireless technologies. The paper at hand intends to fill this research gap with special emphasis on the stochastic analysis of interference dynamics.

\subsection{Contributions}
We derive expressions for the expected value, variance, covariance, and temporal correlation of the interference power at an arbitrary point in space in a network with carrier sensing modeled by an MPP of type~II. Wireless links are modeled by a distance-dependent path loss and Nakagami small-scale fading. The derived expressions are analyzed to highlight the pitfalls of the popular Poisson approximation to CSMA and to explore the impact of system parameters on the correlation of interference. The main novelty is that {\bf expressions for higher-order moments and correlation of interference have been unknown so far for Mat\'ern networks}. These measures show as to how interference changes over time: Is it changing strongly or weakly (variance)? Is it changing quickly or slowly (correlation)? Along these lines, our results give a deeper and more realistic understanding of the {\it temporal dynamics of interference\/} in wireless networks. 

From a practical perspective, such understanding is useful in the design of communication protocols, scheduling, and interference management for many popular wireless technologies for which only rough approximations are available today.
A cooperative relaying protocol, for example, should take into account the correlation of interference to choose its relay.
For instance, a higher correlation requires a higher number of potential relays~\cite{crismani14:tvt}.
Furthermore, LTE Release 13 specifies a licensed-assisted access (LAA) operation mode, that includes a listen-before-talk channel sensing mechanism to support coexistence with WiFi in the unlicensed spectrum (see \cite{feas_study,7782659}). LAA is also perceived to be a key feature of 5G networks~\cite{7497766}.

Our scientific contributions can be summarized as follows: 
\begin{itemize}
\item We derive expressions for the probabilities of one or two points in a PPP being retained in one or two independent Mat\'ern thinnings of the PPP.
\item We derive expressions for the expected value, variance, covariance, and temporal correlation of interference in Mat\'ern networks.
\item We show that interference correlation in Mat\'ern networks is significantly different from that in Poisson networks, which leads to the qualitative conclusion that a PPP is unsuited to capture second-order properties of CSMA networks.
\item We analyze the dependence of the interference correlation on the system parameters and show that, in contrast to  Poisson networks, it depends on the intensity of the MPP and the path loss exponent. Furthermore, fading has a stronger impact on the correlation in networks with carrier sensing.
\end{itemize}

\subsection{Related Work}
Related work mainly includes publications on the modeling of CSMA networks by means of stochastic point processes and publications on interference dynamics in wireless networks. Let us first revisit some work on the use of MPPs in the analysis of interference. The mean interference experienced by a node can be calculated by numerical integration~\cite{haenggi11:mean-interf}; also an upper bound in terms of a closed-form expression is available~\cite{6574907}. Considering network performance, a coverage analysis using a PPP approximation is done in~\cite{baccelli09:vol2}, which is then used to calculate the sensing sensitivity that maximizes the density of concurrent transmissions. Furthermore, the mean throughput~\cite{4215725} and the capacity in case of high SIR~\cite{5752425} have been calculated. All these results have in common that they only consider first-order statistics to analyze the network, i.e., they analyze interference at a single point in time and space. Hence, these results are incapable of capture the dynamic change of interference.
In contrast, we aim to quantify the dynamics of interference (in terms of temporal correlation), which is essential to evaluate a wide range of communication methods, e.g., diversity techniques~\cite{haenggi13:div-poly} and MIMO in millimeter wave communications~\cite{7593259}.

Networks with carrier sensing are also modeled with soft-core point processes~\cite{6841639,6503831,torrisi:ginibre}. Unlike hard-core point processes, there is no strict minimum distance between senders. Instead, a repelling force between senders is assumed that is stronger if they are closer. Hence, it is unlikely for two senders to be very close to each other, yet it is not impossible. This resembles a form of imperfect carrier sensing~\cite{6841639}. Although soft-core processes are probably even more realistic models for CSMA networks, they are rarely applied due to the lack of analytical results.

Alternative models for CSMA networks include perturbed triangular lattices~\cite{7881184} and SSI processes~\cite{busson09:ssi}. These alternatives are investigated almost exclusively by simulations with almost no analytical results available. 

The temporal and spatial correlation of interference is studied in \cite{ganti09:interf-correl,schilcher12:interfcor,tanbourgi14:mrc,crismani14:tvt} but with restriction to~PPPs. 

\smallskip 

The paper is structured as follows: Section~2 explains the modeling assumptions, including node placement, wireless channel, and interference. Section~3 derives expressions for the mean value, variance, covariance, and correlation of interference under this model. Section~4 analyzes the interference correlation and compares results to those of a Poisson network. Section~5 concludes.

\section{Network Model}
\subsection{Spatial Distribution of Senders}

The potential senders in a wireless network are distributed according to a PPP $\ppp\subset\R^2$ with intensity~$\densppp$. Time is partitioned into slots of equal duration. In each slot $t$, some of the potential senders act as senders, i.e., they transmit some data. These senders are modeled by an MPP of type II, denoted by $\tpp\subseteq\ppp$. 
Note that we do not consider the receivers; they are neither included in $\tpp$ nor in $\ppp$. Instead, we assume that each sender has an associated receiver within its range similar to the Poisson bipolar network model (\!\cite{7345601,1580787}).

The decision of a potential sender about sending in a slot $t$ is based on a sensing mechanism for medium access. This mechanism should prevent two nodes from simultaneously sending if their distance is below a certain threshold $\range$, which is similar to CSMA~\cite{baccelli09:vol2}. This behavior is modeled, in each slot, by a dependent thinning of $\ppp$, resulting in an MPP of type~II with intensity $\dens$ for the senders. In other words, the selection of senders is done independently per slot by performing a Mat\'ern thinning of a PPP in each slot.

Such Mat\'ern type II thinning works as follows~\cite{stoyan95}: Each node $x\in\ppp$ draws a uniformly i.i.d. random mark $m_x\sim\uniform$. Node $x$ is retained if and only if $m_x$ is smaller than the marks $m_y$ of all points $y\in \cir(x,\range)$, where $\cir(x,r)=\{y\in\R\,|\,\|x-y\|\leq r\}$ is a circle with radius $r$ centered at~$x$.

The mechanism is fair in the sense that all potential senders are treated equally due to the stationarity of the MPP. It needs to be mentioned that, in real CSMA networks, there is still a possibility that nodes arbitrarily close to each other start sending simultaneously due to imperfect sensing. This effect is not covered by the Mat\'ern model and is outside the scope of this work.

\subsection{Wireless Channel}

The radio propagation is modeled by distance-dependent path loss and small-scale fading caused by multipath propagation. All nodes transmit with unit power. The power $p$ received at the origin $o$ from a sender $x$ is given by $p=\channel{x}$. The term $\ploss{x}=\min(1,\|x\|^{-\alpha})$ is the distance-dependent path gain with exponent $\alpha>2$. The random variable $\fad{x}$ models small-scale fading. We employ the versatile Nakagami-$m$ fading model~\cite{nakagami60:fading}, for which the random variable $\fad{x}$ follows a gamma distribution with parameter $m$, i.e., $\fad{x}\sim\Gamma(m,\frac{1}{m})$ with mean $\expect{\fad{x}}=1$. Recall that the Nakagami model also covers Rayleigh fading ($m=1$) and no fading ($m\to\infty$).

\subsection{Interference}
We are interested in the overall interference at an arbitrary location.
Due to the stationarity of the point processes we consider, without loss of generality, interference at the origin $o$.
This model could be applied for a scenario in which the node suffering from the interference is not an inherent part of the network that causes it. An example application could be mmWave base stations within a 5G cellular network~\cite{7894280} that are interfered by the cellular system without being part of~it.

The interference power at time $t_i$ (time index $i$) is calculated as the sum of the signal powers arriving at the origin $o$ from all active senders, i.e.,
\begin{eqnarray}
\interf_i&=&\sum_{x\in\ppp}\channel{x}\indic{x}{t_i}\\\nonumber
&=&\sum_{x\in\tpp(t_i)}\channel{x}\:,
\end{eqnarray}
where $\indic{x}{t_i}$ is the indicator function that a point $x$ is retained by Mat\'ern thinning at time $t_i$, i.e., that it sends in slot $t_i$.

\section{Interference Expressions}
Our overall goal is to derive the temporal correlation of interference in a given point in space in a Mat\'ern network in terms of Pearson's correlation coefficient $\cor{\interf_1}{\interf_2}$. 
As a basic ingredient for the correlation, we start by calculating the probabilities of a point being retained by the Mat\'ern thinning.
\subsection{Retainment Probabilities}
Let $p_1$ and $p_{12}$ denote the probability that a point is retained once and twice, respectively, and $p_{1/1}(r)$ and $p_{1/2}(r)$ denote the probability that two different points at distance $r$ are retained in one and in two thinnings, respectively.
\begin{lemma}[Single point is retained once]\label{lem:retain:once}
The probability that a point $x\in\ppp$ is retained by Mat\'ern thinning is~\cite{stoyan95}
\begin{equation}\label{eq:pa}
\pa=\frac{1-\exp(-\densppp\,\mc)}{\densppp\,\mc}\:,
\end{equation}
where $\mc$ is the area of a circle with radius $\range$ representing the sensing area of a node.
\end{lemma}
\begin{IEEEproof}[Proof (for didactic purposes)]
For a given mark $m_x$ of $x\in\ppp$ with $0\leq m_x\leq 1$ the point process $\Phi_t=\{y\in\ppp\,|\,m_y<m_x\}$ is an independent $m_x$-thinning of $\ppp$. Hence, it is itself a PPP with intensity $m_x\densppp$. A point $x\in\ppp$ is retained in the Mat\'ern thinning if $\Phi_t\cap b(x,\range)=\emptyset$. Thus, the retainment probability for a given $m_x$ is the void probability $\exp(-m_x\densppp\,\mc)$. Since the mark $m_x$ is chosen by $x$ uniformly in $[0, 1]$, the probability that $x$ is retained in the Mat\'ern thinning is
\begin{equation}
\pa=\int_0^1\exp(-m_x\densppp\,\mc)\,\dd m_x\:.
\end{equation}
Solving this integration yields the result.
\end{IEEEproof}
\textbf{Remarks:}
\begin{itemize}
\item From this Lemma it immediately follows that~\cite{stoyan95}
\begin{equation}\label{eq:dens}
\dens=\densppp\,\pa=\frac{1-\exp(-\densppp\,\mc)}{\mc}\:.
\end{equation}
\item The theoretical maximum intensity of an MPP for a given $\range$ is $\lim_{\densppp\to\infty}\dens=\frac{1}{\range^2\pi}$. Every node possesses an empty guard area of $\range^2\pi$. This point process is sometimes criticized for having low intensities~\cite{busson09:ssi}, but it can actually achieve $\dens$ up to this limit if one chooses a large enough PPP intensity $\densppp$.
\item For $\range\to\infty$ the probability that a point is retained vanishes, i.e., $\lim_{\range\to\infty}\pa=0$.
For $\range\to 0$ all points are retained, i.e., $\lim_{\range\to 0}\pa=1$.
\end{itemize}

\begin{lemma}[Single point is retained twice]\label{lem:pab}
The probability that a point $x\in\ppp$ is retained twice by two independent Mat\'ern thinnings is
\begin{equation}\label{eq:pab}
\pab=\frac{\exp(-\dens\,\mc)\big(\Ei{\dens\,\mc}-\log(\dens\,\mc)-\eulergamma\big)}{\dens\,\mc}\:,
\end{equation}
where $\Ei{x}$ denotes the exponential integral function and $\eulergamma\approx 0.577216$ denotes Euler's $\gamma$ constant. 
\end{lemma}
\begin{IEEEproof}
This proof goes along the lines of the proof of Lemma~\ref{lem:retain:once}.
Let $m_{x,1}$ and $m_{x,2}$ with $0\leq m_{x,1},m_{x,2}\leq 1$ denote the marks of $x\in\ppp$ in the first and the second thinning, respectively. 
We consider the point process $\Phi_{t^2}=\{y\in\ppp\,|\,m_{y,1}<m_{x,1}\vee m_{y,2}<m_{x,2}\}$ of all points having a mark being smaller than $m_{x,1}$ in the first or smaller than $m_{x,2}$ in the second thinning. The probability that an arbitrary point is in this set is $m_{x,1}+m_{x,2}-m_{x,1}m_{x,2}$ by the inclusion-exclusion principle. Hence, the point process $\Phi_{t^2}$ has the intensity $(m_{x,1}+m_{x,2}-m_{x,1}m_{x,2})\,\dens$.
The probability that no point of $\Phi_{t^2}$ is located in $b(x,\range)$ is then given by the void probability of the process $\Phi_{t^2}$, i.e., by $\exp\big(-(m_{x,1}+m_{x,2}-m_{x,1}m_{x,2})\dens\,\mc\big)$. Therefore, the probability that $x$ is retained twice is
\begin{equation}
\pab\!=\!\int_0^1\!\int_0^1\!e^{-(m_{x,1}+m_{x,2}-m_{x,1}m_{x,2})\,\dens\mc} \dd m_{x,1}\,\dd m_{x,2}\:.
\end{equation}
Solving these integrals yields the result.
\end{IEEEproof}
\textbf{Remarks:}
\begin{itemize}
\item The probability $\pab$ that a given point $x$ is retained twice is not the retaining probability squared, i.e., $\pab\neq\pa^2$ for all $\range>0$. The reason is that the number of neighboring points that are potential ``killers'' of $x$ is different for any $x\in\ppp$ but stays constant over time. Hence, the retainings of $x$ in different thinnings are correlated.
\item We have $\pab>\pa^2$ for all $\range>0$, i.e., a point that is retained once is more likely to be retained a second time. This fact can be explained by the following intuition: A point $x$ that is retained once has, on average, fewer neighboring points that could have potentially killed $x$. Therefore, in another independent thinning it has higher chances of being retained.
In the limit, the probabilities converge to $\lim_{\range\to\infty}\pa=\lim_{\range\to\infty}\pab=0$.
\item The intensity of points that are retained twice is given by~$\densppp\,\pab$.
\item For $\range\to 0$ all points are retained twice in two independent thinnings, i.e., $\lim_{\range\to 0}\pab=1$.
\end{itemize}

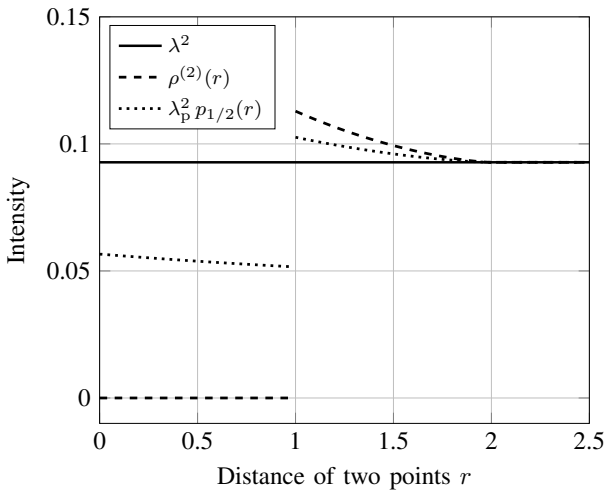
\begin{figure}[b!]
\centering
\begin{tikzpicture}[scale=0.95]
\begin{axis}[xlabel={Distance of two points $r$},ylabel={Intensity},ymin=-0.01,ymax=0.15,xmin=0,xmax=2.5,grid=both,xtick={0,0.5,1,1.5,2,2.5},
            ytick={0,0.05,0.1,0.15,0.2},
xticklabels={0,0.5,1,1.5,2,2.5},
yticklabels={0,0.05,0.1,0.15,0.2},
legend style={at={(0.02,0.98)}, anchor=north west, font=\footnotesize},
legend cell align=left
]

\addplot plot[color=black,solid,no marks,mark=o,line width=1]  table[x index=0,y index=1]  {densities.txt};\addlegendentry{$\dens^2$}
\addplot plot[color=black,dashed,no marks,mark=diamond,line width=1.1,restrict x to domain=0:0.99]  table[x index=0,y index=2]  {densities.txt};\addlegendentry{$\prdensii{r}$}
\addplot plot[color=black,dotted,no marks,mark=triangle,line width=1.1,restrict x to domain=1:2.5]  table[x index=0,y index=3]  {densities.txt};\addlegendentry{$\densppp^2\,\papb(r)$}
\addplot plot[color=black,dashed,no marks,mark=diamond,line width=1.1,restrict x to domain=1:2.5]  table[x index=0,y index=2]  {densities.txt};
\addplot plot[color=black,dotted,no marks,mark=triangle,line width=1.1,restrict x to domain=0:0.99]  table[x index=0,y index=3]  {densities.txt};

\end{axis}
\end{tikzpicture}
\caption{Intensity of the MPP $\dens^2$ (not a function of $r$), of two points separated by $r$ retained in the same thinning $\prdensii{r}$ (the second-order product density), and of two points separated by $r$ each retained in an independent thinning $\densppp^2\,\papb(r)$. Parameters are $\densppp=1$ and $\range=1$.}\label{fig:densities}
\end{figure}

\begin{lemma}[Two distinct points are retained in one thinning]\label{lem:papa}
The probability that two points separated by a distance $r$ are both retained in a Mat\'ern thinning is given by~\cite{stoyan95}
\begin{eqnarray}\label{eq:lem:papa}
\papa(r)&=&\frac{2\Gh\big(1-\exp(-\dens\,\mc)\big)}{\dens^2\mc\Gh\big(\Gh-\mc\big)}\\\nonumber
&&-\,\frac{2\big(1-\exp(-\dens\Gh)\big)}{\dens^2\Gh\big(\Gh-\mc\big)}
\end{eqnarray}
for $r>\range$ and $0$ otherwise. Here, $\Gh$ is the area covered by two overlapping circles with radius $\range$ and centers separated by~$r$, which is given by
\begin{equation}
\Gh=2\mc-\gh\:.
\end{equation}
The overlapping area of these two circles is
\begin{equation}
\gh=2\range^2\arccos\left(\frac{r}{2\range}\right)-\frac{r}{2}\sqrt{4\range^2-r^2}
\end{equation}
for $r\leq 2\range$ and $0$ otherwise~\cite{stoyan95}.
\end{lemma}
\begin{IEEEproof}
Let us consider two points $x,y\in\ppp$ at distance $\|x-y\|=r>0$.
If $r\leq d$, it is impossible that both points are retained due to the definition of a hard-core point process. Hence, let $r>d$ in the following.
Recall that whether $x$ and $y$ are retained depends on the marks of the points in $\cir(x,\range)$ and $\cir(y,\range)$, respectively.
If $r<2d$, these circles overlap and we subdivide them into three areas: $\areac\coloneqq \cir(x,\range)\cap \cir(y,\range)$ is the common area, $\areax\coloneqq \cir(x,\range)\backslash \cir(y,\range)$ and $\areay\coloneqq \cir(y,\range)\backslash \cir(x,\range)$ are the non-common areas. The sizes of these areas are~\cite{stoyan95}
\begin{eqnarray}
|\areac|&=&\gh\,\stackrel{r<2\range}{=}\,2\range^2\arccos\left(\frac{r}{2\range}\right)-\frac{r}{2}\sqrt{4\range^2-r^2}\hspace{5mm}\label{eq:pr:ac}\\
|\areax|&=&|\areay|\,=\,\range^2\pi-\gh\:.\label{eq:pr:axy}
\end{eqnarray}
For $r\geq 2d$ the common area vanishes giving $|\areac|=0$.
The area covered by at least one of the circles is
\begin{equation}
|\areax\cup\areay|=\Gh=2\range^2\pi-\gh\:.
\end{equation}

Let $m_x$ and $m_y$ denote the marks of $x$ and $y$, respectively. To retain both $x$ and $y$, the following three conditions have to hold: Firstly, $\areax$ must not contain any point from $z\in\ppp$ with $m_z<m_x$. For given $m_x$, the probability for it is $\exp(-m_x\, \densppp\, |\areax|)$, similarly to the proofs of Lemma~\ref{lem:retain:once} and~\ref{lem:pab}.
Secondly, $\areay$ must not contain any $z\in\ppp$ with $m_z<m_y$, which happens with probability $\exp(-m_y\, \densppp\, |\areay|)$. 
Thirdly, the common area $\areac$ must not contain any $z\in\ppp$ with $m_z<\max(m_x,m_y)$, which has the probability $\exp(-\max(m_x,m_y)\,\densppp\, |\areac|)$.
Overall, the probability that both $x$ and $y$ are retained~is
\begin{eqnarray}
\papa(r)&=&\int_0^1\int_0^1\exp\big(-(m_x+m_y)\,\densppp\,|\areax|\\\nonumber
&&-\max(m_x,m_y)\,\densppp\,|\areac|\big)\,\dd m_x\,\dd m_y\:.
\end{eqnarray}
Solving these integrals yields the result.
\end{IEEEproof}
\textbf{Remarks:}
\begin{itemize}
\item Based on the probability in~\eqref{eq:lem:papa}, we can calculate the second-order product density of the MPP by
$\prdensii{r}=\densppp^2\,\papa(r)$. A plot of it is shown in Fig.~\ref{fig:densities}.
\item For the case $r>2\range$ we have $\gh=0$ and hence $\prdensii{r}=\dens^2$, since $\papa(r)\stackrel{r>2\range}{=}\pa^2$. This implies that two points that are further than $2\range$ apart from each other are independently retained or removed. Recall that the second-order product density for PPPs is $\densppp^2$.
\item For the case $\range<r\leq 2\range$ the integration over $\prdensii{r}$ yields no closed form solution due to the complexity of $\gh$. Hence, in some cases it might be advantageous to approximate it by $\gh\approx\range^2\pi-2\range r$~\cite{stoyan95}.
\end{itemize}
\begin{figure*}
	\begin{eqnarray}\label{eq:papb}
	  \papb(r)&\stackrel{r\geq\range}{=}&\frac{\exp\left(-\frac{\range^4\pi^2\densppp}{\gh}\right)\left(2\Gamma\left(0,\mc\densppp\left(1-\frac{\mc}{\gh}\right)\right)-\Gamma\left(0,-\frac{\range^4\pi^2\densppp}{\gh}\right)-\Gamma\left(0,-\frac{\densppp\big(\gh-\mc\big)^2}{\gh}\right)\right)}{\densppp\,\gh}\\\label{eq:papbii}
	\label{eq:papbcl}
	\papb(r)&\stackrel{r<\range}{=}&\frac{\exp\left(-\range^2\pi\densppp\left(2+\frac{\range^2\pi}{\gh}\right)\right)}{\range^2\pi\densppp^2\ghpot{3}}
	\vast(\exp(2\range^2\pi\densppp)\Big(\gh\big(1+2\range^2\pi\densppp-\gh\densppp\big)-\range^4\pi^2\densppp\Big)\\\nonumber
	&&\nvast(2\Ei{\range^2\pi\densppp\left(\frac{\range^2\pi}{\gh}-1\right)}-\Ei{\frac{\range^4\pi^2\densppp}{\gh}}-
	\Ei{\frac{\densppp\big(\gh-\range^2\pi\big)^2}{\gh}}\nvast)\\\nonumber
	&&-\gh\exp\left(\frac{\range^4\pi^2\densppp}{\gh}\right)\Big(\exp(\densppp\gh)\range^2\pi+\exp(2\range^2\pi\densppp)(\range^2\pi-2\gh)-2\exp(\range^2\pi\densppp)\big(\range^2\pi-\gh\big)\Big)\vast)
	\end{eqnarray}
	\hrule
\end{figure*}
\begin{lemma}[Two distinct points are retained in independent thinnings]\label{lem:papb}
The probability $\papb(r)$ that a point $x\in\ppp$ is retained by a thinning and $y\in\ppp$ is retained in another, independent thinning with $r=\|x-y\|>0$ is given in~\eqref{eq:papb} for $r\geq\range$ and in~\eqref{eq:papbcl} for $r<\range$.
\end{lemma}
\begin{IEEEproof}
Let us consider two points $x,y\in\ppp$ at distance $\|x-y\|=r>0$.
We define the areas $\areac$, $\areax$ and $\areay$ as given in~\eqref{eq:pr:ac} and~\eqref{eq:pr:axy} of the proof of Lemma~\ref{lem:papa}.

Let us assume $r>\range$.
For given $m_x$ and $m_y$, there should be no point $z\in\ppp$ with $m_z<m_x$ in the area $\areax$ at the first thinning and no point $z\in\ppp$ with $m_z<m_y$ in $\areay$ at the second thinning. The probability for these events is given by
\begin{equation}
	\exp\Big(-(m_x+m_y)\,\densppp\,\big(\mc-\gh\big)\Big)\:.
\end{equation}
The probability that there is no point $z\in\ppp$ in $\areac$ with $m_z<m_x$ at the first or $m_z<m_y$ at the second thinning~is
\begin{equation}
\exp\big(-(m_x+m_y-m_xm_y)\,\densppp\,\gh\big)
\end{equation}
similar to the proof of Lemma~\ref{lem:pab}.
Integrating over the product of these two expressions yields
\begin{eqnarray} \label{eq:pr:papbl}
\papb(r)&\hspace{-3mm}\stackrel{r>\range}{=}\hspace{-3mm}&\int_0^1\hspace{-1mm}\int_0^1\exp\Big(\hspace{-1mm}-(m_x+m_y)\,\densppp\,\big(\mc-\gh\big)\hspace{5mm}\\  \nonumber
&&-\,(m_x+m_y-m_xm_y)\,\densppp\,\gh\Big)\,\dd m_x\,\dd m_y\:. 
\end{eqnarray}

Next, we assume that $r<\range$. In this case, the derivation of $\papb(r)$ is similar to the previous case, except $x\in\areac$ in the first thinning and $y\in\areac$ in the second one. Hence, $x$ has to have a higher mark than $y$ in the first thinning, and $y$ a higher mark than $x$ in the second thinning. These two events have the probabilities $(1-m_x)$ and $(1-m_y)$ respectively, leading to
\begin{eqnarray}\label{eq:pr:papbs}
\papb(r)&\stackrel{r\leq\range}{=} \nonumber &\int_0^1\int_0^1\exp\Big(-(m_x+m_y)\,\densppp\,\big(\mc-\gh\big)\\ 
&&-\,(m_x+m_y-m_xm_y)\,\densppp\,\gh\Big)\\\nonumber
&&(1-m_x)(1-m_y)\,\dd m_x\,\dd m_y\:.
\end{eqnarray}
Solving the integrations in~\eqref{eq:pr:papbl} and~\eqref{eq:pr:papbs} yields the result.
\end{IEEEproof}

\textbf{Remarks:}
\begin{itemize}
\item The probability $\papa(r)$ vanishes for $r<\range$ since two points closer than $\range$ cannot both be retained in one thinning. However, the probability that each of them is retained in an independent thinning is $\papb(r)>0$ for $r<\range$. Still, it is much smaller than for $r\geq\range$ (see Fig.~\ref{fig:densities}), since for $r\geq\range$ they are neighbors and could potentially kill each other, i.e., the random number of neighbors is higher by one in this case. 
\item If the two points approach each other and become identical, the probability $\papb(r)$ becomes the probability $\pab$ that one point is retained twice, i.e., $\lim_{r\to 0}\papb(r)=\pab$. Care has to be taken when calculating this limit: We have to adopt~\eqref{eq:papb}, which is intended for $r\geq\range$ and does not consider an extra point in the neighborhood. Hence, in the limit when the two points merge to become one, there is no extra point in the neighborhood leading to~\eqref{eq:pab}. Calculating the limit of~\eqref{eq:papbcl} leads to the different expression $\frac{\exp(-\dens\,\mc)(1+E_i(\dens\,\mc)-\log(\dens\,\mc)-\eulergamma)-1}{\dens\,\mc}$.
\item If $r>2\range$, the events that two points are retained each in an independent thinning are independent. Therefore, we have $\papb(r)\stackrel{r>2\range}{=}\pa^2$.
\end{itemize}

Note that for $\range\to 0$ all results in this section converge to the corresponding results of a PPP.

\subsection{Expected Value, Variance, and Covariance of Interference}
\begin{lemma}[Expected interference]
The expected value of interference is
\begin{equation}\label{eq:expectedinterference}
\expect{\interf}=\dens\,\frac{\alpha\pi}{\alpha-2}\:.
\end{equation}
\end{lemma}
\begin{IEEEproof}
All nodes in the set $\tpp\subseteq\ppp$ are considered to be interferers.
Hence, the expected value of interference is calculated by applying Campbell's theorem yielding
\begin{eqnarray}
\expect{\interf}
&=&\expect{\sum_{x\in\ppp}\channel{x}\indic{x}{t}}\\\nonumber
&=&\densppp\,\int_{\R^2}\ploss{x}\,\expect{\fad{x}}\,\expect{\indic{x}{t}}\,\dd x\\\nonumber
&=&\densppp\pa\,\frac{\alpha\pi}{\alpha-2}\:.
\end{eqnarray}
The expected value of the indicator function $\indic{x}{t}$ does neither depend on $x$ nor on $t$. This is because we do not consider any nodes to be placed at certain locations (e.g., the origin), which implies that we do not adopt the Palm distribution of the MPP.
Substituting~\eqref{eq:dens} into this expression gives the result.
\end{IEEEproof}
{\bf Remarks:}
\begin{itemize}
\item The expected interference \eqref{eq:expectedinterference} for MPP is the same as the one for PPP (given in \cite{haenggi13:book,schilcher12:tmc}) with the same process intensity using the same path loss model. 

\item The expression \eqref{eq:expectedinterference} does not correspond to the expected interference derived by Haenggi for MPP~\cite{haenggi11:mean-interf}. The core difference is in the modeling assumptions: Haenggi assumes that a node is located at the origin; his result thus represents the expected interference experienced by a typical node in the network. This assumption causes some mathematical difficulties, as the reduced Palm distribution has to be adopted. We do not assume a node to be located at the origin; all results hold for any point in space. Nevertheless, a data sink could be located at this point that neither sends data nor participates in the CSMA protocol. Along these lines, our assumptions are well-suited for a multipoint-to-point communication~scenario. 
\end{itemize}

\pagebreak
\begin{theorem}[Variance of interference]\label{th:var}
The variance of interference at the origin $o$ is
\begin{eqnarray}\label{eq:var}
\var{\interf}&=&\dens\,\frac{(m+1)\,\alpha\pi}{m\,(\alpha-1)}\\\nonumber
&&+\,8\pi\int_\frac{\range}{2}^\infty\int_0^\infty\int_0^{2\pi}\plosssc{r(\cosh\mu+\cos\nu}\\\nonumber
&&\plosssc{r(\cosh\mu-\cos\nu}\frac{r^2}{2}\big(\cosh 2\mu-\cos 2\nu\big)\,\dd \nu\,\dd \mu\,\\\nonumber
&&\prdensii{2r}\,r\,\dd r-\left(\frac{\dens\alpha\pi}{\alpha-2}\right)^2\:,
\end{eqnarray}
where $\dens$ is the intensity of the MPP, $m$ is the parameter of Nakagami fading, $\alpha$ is the path loss exponent, $\range$ is the hard-core distance, and $\plosssc{\cdot}$ is the path gain function.
\end{theorem}
\begin{IEEEproof}
We start by calculating the second moment of interference at time $t$ by
\begin{eqnarray}\label{eq:pr:I2sep}
\lefteqn{\expect{\interf^2}=}\\\nonumber
&=&\mathbb{E}\nvast[\left(\sum_{x\in\ppp}\channel{x}\indic{x}{t}\right)\\\nonumber
&&\hspace{1cm}\cdot\,\left(\sum_{y\in\ppp}\channel{y}\indic{y}{t}\right)\nvast]\\\nonumber
&\stackrel{(a)}{=}&\expect{\sum_{x\in\ppp}\big(\channel{x}\big)^2\indic{x}{t}}\\\nonumber
&&+\,\expect{\sum_{x,y\in\ppp}^{\neq}\channel{x}\channel{y}\indic{x}{t}\indic{y}{t}}\:,
\end{eqnarray}
where in $(a)$ terms with $x=y$ are separated from terms with $x\neq y$.
The first of these expected values yields
\begin{eqnarray}
\lefteqn{\expect{\sum_{x\in\ppp}\big(\channel{x}\big)^2\indic{x}{t}}=}\\\nonumber
&\stackrel{(a)}{=}&\densppp\pa\,\expect{\fadsq{x}}\int_{\R^2}\plosssq{x}\,\dd x\\\nonumber
&=&\dens\,\frac{(m+1)\,\alpha\pi}{m\,(\alpha-1)}\:,
\end{eqnarray}
where in $(a)$ we apply Campbell's theorem.
The last expectation of~\eqref{eq:pr:I2sep} gives
\begin{eqnarray}\label{eq:pr:I2neq}
\lefteqn{\expect{\sum_{x,y\in\ppp}^{\neq}\channel{x}\channel{y}\indic{x}{t}\indic{y}{t}}=}\\\nonumber
&\stackrel{(a)}{=}&\int_{\R^2}\int_{\R^2}\ploss{x}\,\ploss{y}\,\prdensii{\|x-y\|}\,\dd x\,\dd y\\\nonumber
&\stackrel{(b)}{=}&4\int_{\R^2}\int_{\R^2}\ploss{x}\,\ploss{x-2a}\,\prdensii{\|2a\|}\,\dd x\,\dd a\\\nonumber
&\stackrel{(c)}{=}&4\int_{\R^2}\int_{\R^2}\ploss{x+a}\,\ploss{x-a}\,\dd x\,\prdensii{\|2a\|}\,\dd a\\\nonumber
&\stackrel{(d)}{=}&8\pi\int_\frac{\range}{2}^\infty\int_{\R^2}\ploss{x+\begin{pmatrix}r\\0\end{pmatrix}}\,\ploss{x-\begin{pmatrix}r\\0\end{pmatrix}}\,\dd x\,\\\nonumber
&&\prdensii{2r}\,r\,\dd r\:.
\end{eqnarray}
In $(a)$ we apply a basic property of the second-order product density $\prdensii{r}=\densppp^2\,\papa(r)$~\cite[p.~112]{stoyan95}, where $\papa(r)$ is the probability that two points at distance $r$ are both retained, as derived in Lemma~\ref{lem:papa}. Furthermore, the expected values of the fading coefficients $\expect{\fad{z}}=1$ for any $z\in\ppp$ are substituted. In $(b)$ we substitute $y=x-2a$, and $4$ is the corresponding Jacobi determinant; in $(c)$ we substitute $x+a$ for $x$.
In $(d)$ we substitute polar coordinates. The integration of $r$ starts at $\frac{\range}{2}$ as $\prdensii{r}=0$ for $r<\range$. Furthermore, we apply a rotation of the coordinate system to translate $a$ into a real number $a'$:
\begin{equation}
{\bf x'}=\left(
	\begin{array}{cc}
	\cos\phi  & \sin\phi \\
	-\sin\phi & \cos\phi
	\end{array}\right){\bf x},
\end{equation}
where ${\bf x'}$ and ${\bf x}$ are the vector notations of points $x'$ and $x$ in $\R^2$, respectively, and $\phi = \arctan(a_I/a_R)$ is the phase of the polar coordinate of $a$. This transformation does not change the norms $\ploss{x+a}$, $\ploss{x-a}$, and $\ploss{2a}$ involved in the integration.
This can be verified by writing $x'=e^{j\phi}x$ and $a'=e^{j\phi}a$. 

For an arbitrary but fixed $r\in\R$, the inner integral of~\eqref{eq:pr:I2neq} is
\begin{eqnarray}
\lefteqn{\int_{\R^2}\ploss{x+a}\,\ploss{x-a}\,\dd x=}\\\nonumber
&\stackrel{(a)}{=}&\int_0^\infty\int_0^{2\pi}\plosssc{r(\cosh\mu+\cos\nu}\,\\\nonumber
&&\plosssc{r(\cosh\mu-\cos\nu}\frac{r^2}{2}\big(\cosh 2\mu-\cos 2\nu\big)\,\dd \nu\,\dd \mu\:.
\end{eqnarray}
In $(a)$ we substitute elliptic coordinates defined as 
\begin{eqnarray}
x_1&=&r\,\cosh\mu\,\cos\nu\\\nonumber
x_2&=&r\,\sinh\mu\,\sin\nu
\end{eqnarray}
and its corresponding Jacobi determinant $\frac{r^2}{2}\big(\cosh 2\mu-\cos 2\nu\big)$, and calculate the corresponding norms.
Calculating $\var{\interf}=\expect{\interf^2}-\expect{\interf}^2$ yields the result.
\end{IEEEproof}
\textbf{Remarks:}
\begin{itemize}
\item The variance of interference is presented in terms of integral expressions that are solved numerically as there is no closed-form solution. The problematic term for symbolic integration is the second-order product density~$\prdensii{r}$.
\item Since numerical integration is involved in calculating the variance of interference some steps of the proof would not be needed. For example, the expression in step $(a)$ of~\eqref{eq:pr:I2neq} could directly be solved numerically. However, it turns out that it is advantageous to rather solve the integration in~\eqref{eq:var} as it results in better numerical stability.
\item The substitution of elliptic coordinates might be an interesting approach for other applications: It allows to solve integrations of the form $\int_{\R^2}\|x-a\|\|x+a\|\,\dd x$, which sometimes occur in the derivation of second order statistics of interference for both PPP and MPP.
\item When $\range\to 0$ the variance converges to the Poisson case: In~\eqref{eq:pr:I2neq} $(a)$ the second-order product density $\prdensii{r}$ could be substituted by $\dens^2$ leading to
\begin{eqnarray}
\lefteqn{\expect{\sum_{x,y\in\ppp}^{\neq}\channel{x}\channel{y}\indic{x}{t}\indic{y}{t}}=}\\\nonumber
&=&\left(\dens\int_{\R^2}\ploss{x}\,\dd x\right)^2\\\nonumber
&=&\left(\dens\,\frac{\alpha\pi}{\alpha-2}\right)^2\:.\hspace{3.5cm}
\end{eqnarray}
Since this expression is equal to $\expect{\interf}^2$, we have
\begin{equation}
\var{\interf}=\dens\,\frac{(m+1)\,\alpha\pi}{m\,(\alpha-1)}
\end{equation}
equal to the Poisson case~\cite{ganti09:interf-correl}.
\end{itemize}

\begin{theorem}[Covariance of interference]\label{th:cov}
The temporal covariance of interference at the origin $o$ is
\begin{eqnarray}\label{eq:cov}
\lefteqn{\cov{\interf_1}{\interf_2}=\densppp\pab\,\frac{\alpha\pi}{\alpha-1}}\\\nonumber
&&+\,8\pi\int_0^\infty\int_0^\infty\int_0^{2\pi}\plosssc{r(\cosh\mu+\cos\nu}\\\nonumber
&&\plosssc{r(\cosh\mu-\cos\nu}\frac{r^2}{2}\big(\cosh 2\mu-\cos 2\nu\big)\,\dd \nu\,\dd \mu \\\nonumber
&&\papb(2r)\,r\,\dd r-\left(\frac{\dens\alpha\pi}{\alpha-2}\right)^2\:,
\end{eqnarray}
where $\densppp$ is the intensity of the PPP, $\dens=\densppp\pa$ is the intensity of the MPP, $m$ is the parameter of Nakagami fading, $\alpha$ is the path loss exponent, $\range$ is the hard-core distance, and $\plosssc{\cdot}$ is the path gain function. $\papb(r)$ denotes the probability that two different nodes are retained in two independent Mat\'ern thinnings, as derived in Lemma~\ref{lem:papb}.
\end{theorem}
\begin{IEEEproof}
The proof goes along the lines of the proof of Theorem~\ref{th:var}.
We start by calculating the covariance of interference at times $t_1$ and $t_2$ by
\begin{eqnarray}\label{eq:pr:I1I2sep}
\lefteqn{\expect{\interf_1\interf_2}=}\\\nonumber
&=&\mathbb{E}\nvast[\left(\sum_{x\in\ppp}\channel{x}\indic{x}{t_1}\right)\\\nonumber
&&\hspace{1cm}\cdot\,\left(\sum_{y\in\ppp}\channel{y}\indic{y}{t_2}\right)\nvast]\\\nonumber
&\stackrel{(a)}{=}&\expect{\sum_{x\in\ppp}\plosssq{x}\indic{x}{t_1}\indic{x}{t_2}}\\\nonumber
&&+\,\expect{\sum_{x,y\in\ppp}^{\neq}\ploss{x}\ploss{y}\indic{x}{t_1}\indic{y}{t_2}}\:,
\end{eqnarray}
where in $(a)$ terms with $x=y$ are separated from terms with $x\neq y$ and the expected value of fading $\expect{\fad{x}}=1$ is substituted.
The first of these expected values yields
\begin{eqnarray}
\lefteqn{\expect{\sum_{x\in\ppp}\plosssq{x}\indic{x}{t_1}\indic{x}{t_2}}=}\\\nonumber
&\stackrel{(a)}{=}&\densppp\pab\,\int_{\R^2}\plosssq{x}\,\dd x\\\nonumber
&=&\densppp\pab\,\frac{\alpha\pi}{\alpha-1}\:,
\end{eqnarray}
where in $(a)$ we apply Campbell's theorem.
The second expectation of~\eqref{eq:pr:I1I2sep} gives
\begin{eqnarray}\label{eq:pr:I1I2neq}
\lefteqn{\expect{\sum_{x,y\in\ppp}^{\neq}\channel{x}\channel{y}\indic{x}{t_1}\indic{y}{t_2}}=}\\\nonumber
&\stackrel{(a)}{=}&\densppp^2\int_{\R^2}\int_{\R^2}\ploss{x}\,\ploss{y}\,\papb(\|x-y\|)\,\dd x\,\dd y\:.\hspace{5mm}
\end{eqnarray}
In $(a)$ we apply a basic property of the process, where $\papb(r)$ is the probability that two points at distance $r$ are both retained each in an independent Mat\'ern thinning, as derived in Lemma~\ref{lem:papb}. Note that this expression is similar to~\eqref{eq:pr:I2sep} except that the second-order product density $\prdensii{2r}$ is substituted by $\densppp^2\,\papb(2r)$.
Applying similar steps as in the proof of Theorem~\ref{th:var} yields the~result.
\end{IEEEproof}
\textbf{Remarks:}
\begin{itemize}
\item The covariance does not depend on fading as there is no $m$ in the expression.
\item Again, for $\range\to 0$ the expression of $\cov{\interf_1}{\interf_2}$ converges to the Poisson case: As mentioned in the remarks after Lemma~\ref{lem:papb}, we have $\papb(r)\stackrel{r>2\range}{=}\pa^2$. Substituting this result into~\eqref{eq:pr:I1I2neq} leads to
\begin{eqnarray}
\lefteqn{\expect{\sum_{x,y\in\ppp}^{\neq}\channel{x}\channel{y}\indic{x}{t_1}\indic{y}{t_2}}=}\\\nonumber
&=&\left(\dens\int_{\R^2}\ploss{x}\,\dd x\right)^2\:,\hspace{3cm}
\end{eqnarray}
which is equal to $\expect{\interf}^2$. Thus, we have
\addtocounter{equation}{1}
\begin{equation}
\cov{\interf_1}{\interf_2}=\densppp\pab\,\frac{\alpha\pi}{\alpha-1}\:.
\end{equation}
\end{itemize}

\subsection{Correlation of Interference}
\begin{corollary}[Correlation of interference]\label{cor:cor}
The temporal correlation of interference $\cor{\interf_1}{\interf_2}$ at the origin $o$ is given in~\eqref{eq:cor}.
\begin{figure*}
	\begin{equation}\label{eq:cor}
\cor{I_1}{I_2}=
\frac{\frac{\dens(m+1)\,\alpha\pi}{m\,(\alpha-1)}+
8\pi\!\int_\frac{\range}{2}^\infty\!\!\int_0^\infty\!\!\int_0^{2\pi}\frac{\plosssc{r(\cosh\mu+\cos\nu}
\plosssc{r(\cosh\mu-\cos\nu}r^2(\cosh 2\mu-\cos 2\nu)}{2}\,\dd \nu\,\dd \mu
\prdensii{2r}r\,\dd r-\left(\frac{\dens\alpha\pi}{\alpha-2}\right)^2}
{\frac{\densppp\pab\alpha\pi}{\alpha-1}+
8\pi\!\int_0^\infty\!\!\int_0^\infty\!\!\int_0^{2\pi}\frac{\plosssc{r(\cosh\mu+\cos\nu}
\plosssc{r(\cosh\mu-\cos\nu}r^2(\cosh 2\mu-\cos 2\nu)}{2}\,\dd \nu\,\dd \mu
\papb(2r)r\,\dd r-\left(\frac{\dens\alpha\pi}{\alpha-2}\right)^2}\:.
	\end{equation}
	\hrule
\end{figure*}
\end{corollary}
\begin{IEEEproof}
Pearson's correlation coefficient is defined as $\cor{\interf_1}{\interf_2}=\frac{\cov{\interf_1}{\interf_2}}{\sqrt{\var{\interf_1}\,\var{\interf_2}}}$.
Hence, the result is obtained by dividing~\eqref{eq:cov} by~\eqref{eq:var}.
\end{IEEEproof}
\textbf{Remarks:}
\begin{itemize}
\item In the limit $\range\to 0$ the temporal correlation of interference approaches the Poisson case with all potential senders transmitting:
\begin{equation}
\lim_{\range\to 0}\cor{I_1}{I_2}=\frac{m}{m+1}\:.
\end{equation}
In particular, for Rayleigh fading ($m=1$) the correlation is $\lim_{\range\to 0}\cor{I_1}{I_2}\stackrel{m=1}{=}\frac{1}{2}$ and without fading we have $\lim_{m\to\infty}\lim_{\range\to 0}\cor{I_1}{I_2}=1$.
\item The temporal correlation of interference does not depend on the time period between $\interf_i$ and $\interf_j$. In other words, for a given time instant $t$, the correlation $\cor{\interf_t}{\interf_{t+i}}$ is the same for all $i\in\mathbb{Z}$. This result is relevant for retransmission protocols: If a transmission failed and has to be repeated, the sender can expect the same interference statistics independent of the time instant of the retransmission, i.e., a longer backoff does not increase the success probability in this model.
\end{itemize}

\section{\!\mbox{Insights on Interference Correlation}}
So far we have derived expressions for the MPP and explained how these are generalizations of the PPP.
Let us now plot and analyze the interference correlation of the Mat\'ern network over certain parameters and compare these results to those of a Poisson network. 
For a fair comparison, the senders in the Poisson network are selected by an independent thinning of the PPP with probability $\pa$, leading to an intensity $\dens$ (which is the density of senders in the Mat\'ern network). This model resembles ALOHA as MAC protocol, where the sending probability is $\pa$. Both the CSMA and the ALOHA network model have the same expected value of interference $\expect{\interf}$. Remember that, unlike for PPPs, conditioning on a point at the origin $o$ does change the distribution of the rest of the process for MPPs, since there cannot be any point in its vicinity, i.e., $\cir(o,\range)\cap\tpp\backslash\{o\}=\emptyset$. In this work, we do not condition on having a point of the process being located at the origin $o$.

If not stated otherwise, we use a path loss exponent $\alpha=3$ and an intensity $\densppp=1$ for the PPP, hence having $0\leq\dens\leq 1$, depending on the value of $\range$. The hard-core distance is $\range=1$. 

All mathematical results have been crosschecked by simulations, showing a good match. All plots show only the mathematical results because the simulation results do not provide any additional information. 

\begin{figure}[th!]
\centering
\begin{tikzpicture}
\begin{axis}[xlabel={Fraction of nodes sending $\frac{\dens}{\densppp}$},ylabel={Interference correlation $\cor{\interf_1}{\interf_2}$},ymin=0,ymax=1.05,xmin=0,xmax=1,grid=both,xtick={0,0.2,0.4,0.6,0.8,1},
            ytick={0,0.2,0.4,0.6,0.8,1.0},
xticklabels={0,0.2,0.4,0.6,0.8,1},
yticklabels={0,0.2,0.4,0.6,0.8,1.0},
legend style={at={(0.02,0.98)}, anchor=north west, font=\footnotesize},
legend cell align=left
]

\addplot plot[color=black,dotted,no marks,style=thick]  table[x index=0,y index=21]  {cor_dens.txt};
\addplot plot[color=black,only marks,mark=x,style=thick]  table[x index=0,y index=0]  {cor_dens.txt};
\addplot plot[color=black,loosely dashed,no marks,style=thick]  table[x index=0,y index=2]  {cor_dens.txt};
\addplot plot[samples=200,color=black,only marks,mark=diamond,style=thick]  (\x,{\x/2});
\addplot plot[color=black,dashdotted,no marks,style=thick]  table[x index=0,y index=1]  {cor_dens.txt};
\addplot plot[samples=200,color=black,only marks,mark=triangle,style=thick]  (\x,{\x/3});

\draw (axis cs: 0.94,0.95) ellipse (2 and 8);
\node at (axis cs: 0.9,0.83) {$m\rightarrow\infty$};

\draw (axis cs: 0.93,0.46) ellipse (2 and 5);
\node at (axis cs: 0.9,0.55) {$m=1$};

\draw (axis cs: 0.93,0.31) ellipse (2 and 6);
\node at (axis cs: 0.9,0.2) {$m=\frac{1}{2}$};

\end{axis}
\end{tikzpicture}
\caption{Temporal correlation of interference over different densities of transmitters. Lines indicate MPP and marks indicate PPP. Parameters are $\densppp=1$ and $\alpha=3$.}\label{fig:dens}
\end{figure}
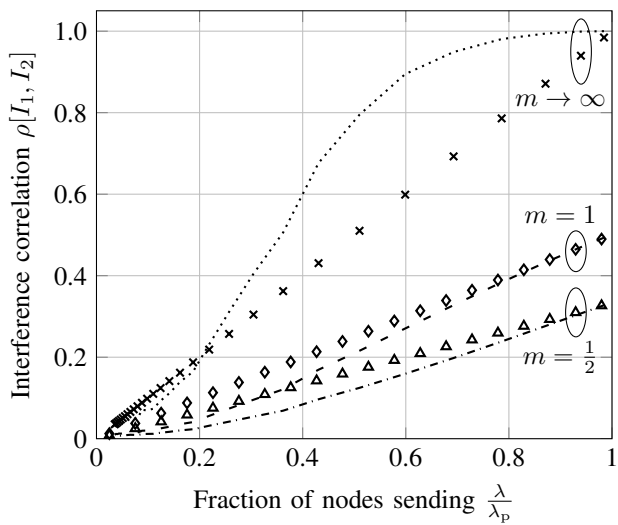

\subsection{Impact of Fraction of Sending Nodes}
Fig.~\ref{fig:dens} shows the temporal correlation of interference for different fractions of nodes acting as interferers for both PPP and MPP. 
For the MPP, this correlation is calculated by numerical integration of the expression in Corollary~\ref{cor:cor} and is plotted over $\pa$ given in~\eqref{eq:pa}, varying the value of $\range$. For the PPP, the scenario corresponds to Case~$(2,1,1)$ in the classification of~\cite{schilcher12:interfcor}; the correlation is $\frac{q m}{m+1}$ with $q$ being the fraction of active senders~\cite{ganti09:interf-correl,schilcher12:interfcor}.

The most apparent observation is that PPP and MPP yield significantly different correlation curves. Hence, adopting a PPP to model a CSMA network may lead to the correct average interference, but it will incorrectly estimate the temporal dynamics in terms of correlation. 
Nevertheless, the curves of both models show the same trend: The correlations are strictly monotonically increasing with the fraction of senders and hit the same maximum value of $\frac{m}{m+1}$ for $\dens\to\densppp$, i.e., $\pa\to 1$. In particular, the interference correlation of a PPP is neither an upper nor a lower bound for the one of an MPP. In general, it is higher for a small fraction of senders and lower for a high fraction. The crossing point heavily depends on the fading: weak fading (high $m$) shifts the crossing point toward small fractions of~senders.

\subsection{Impact of Fading}

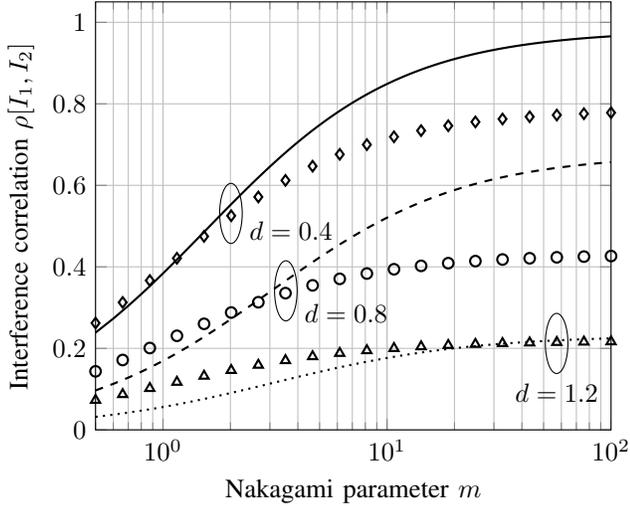
\begin{figure}[t!]
\centering
\begin{tikzpicture}
\begin{axis}[xlabel={Nakagami parameter $m$},ylabel={Interference correlation $\cor{\interf_1}{\interf_2}$},ymin=0,ymax=1.05,xmin=0.5,xmax=100,grid=both,
xmode=log,
legend style={at={(0.02,0.98)}, anchor=north west, font=\footnotesize},
legend cell align=left
]

\addplot plot[color=black,solid,no marks,style=thick]  table[x index=0,y index=4]  {cor_m.txt};
\addplot plot[samples=20,domain=0.5:100,color=black,only marks,mark=diamond,style=thick]  (\x,{0.78598*\x/(\x+1)});

\addplot plot[color=black,dashed,no marks,style=thick]  table[x index=0,y index=8]  {cor_m.txt};
\addplot plot[samples=20,domain=0.5:100,color=black,only marks,mark=o,style=thick]  (\x,{0.43076*\x/(\x+1)});

\addplot plot[color=black,dotted,no marks,style=thick]  table[x index=0,y index=12]  {cor_m.txt};
\addplot plot[samples=20,domain=0.5:100,color=black,only marks,mark=triangle,style=thick]  (\x,{0.21865*\x/(\x+1)});

\end{axis}

\draw (1.8,2.88) ellipse (.15 and .4);
\node at (2.6,2.65) {$\range=0.4$};
\draw (2.53,1.85) ellipse (.15 and .4);
\node at (3.33,1.55) {$\range=0.8$};
\draw (6.13,1.15) ellipse (.15 and .4);
\node at (6.13,.5) {$\range=1.2$};

\end{tikzpicture}
\caption{Temporal correlation of interference over different values of the fading parameter $m$. Lines indicate MPP and marks indicate PPP. Parameters are $\densppp=1$~and~$\alpha=3$.}\label{fig:m}
\end{figure}

Fig.~\ref{fig:m} shows  how the interference correlation depends on the severeness of fading represented by $m$. It is well known for PPPs that the correlation of interference decreases with increasing fading (decreasing $m$). The reason is that severe fading leads to a high variance of interference but does not change the covariance. We observe the same qualitative behavior for MPPs. All curves flatten out for increasing $m$, and in case of no fading ($m\to\infty$), the correlation converges to a value that depends on $\dens$ and $\range$, which is plotted in Fig.~\ref{fig:dens} (dotted curve). The value of $m$ determines as to which of the two models shows a higher correlation. Last but not least, the curves show that severe fading (small $m$) has higher impact on the correlation of MPPs than of PPPs.  

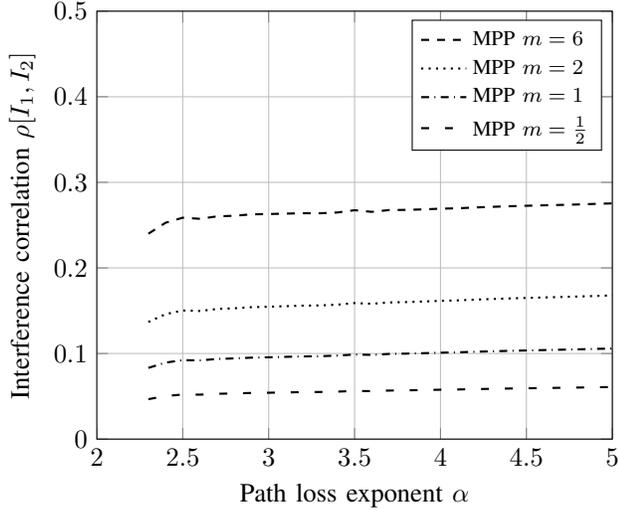
\begin{figure}[t!]
\centering
\begin{tikzpicture}
\begin{axis}[xlabel={Path loss exponent $\alpha$},ylabel={Interference correlation $\cor{\interf_1}{\interf_2}$},ymin=0,ymax=0.5,xmin=2,xmax=5,grid=both,
legend style={at={(0.98,0.98)}, anchor=north east, font=\footnotesize},
legend cell align=left
]

\addplot plot[color=black,dashed,no marks,style=thick]  table[x index=0,y index=12]  {cor_al.txt};\addlegendentry{MPP $m=6$}
\addplot plot[color=black,dotted,no marks,style=thick]  table[x index=0,y index=4]  {cor_al.txt};\addlegendentry{MPP $m=2$}
\addplot plot[color=black,dashdotted,no marks,style=thick]  table[x index=0,y index=2]  {cor_al.txt};\addlegendentry{MPP $m=1$}
\addplot plot[color=black,loosely dashed,no marks,style=thick]  table[x index=0,y index=1]  {cor_al.txt};\addlegendentry{MPP $m=\frac{1}{2}$}

\end{axis}
\end{tikzpicture}
\caption{Temporal correlation of interference over different values of the path loss exponent $\alpha$. Parameters are $\densppp=1$ and $\range=1$. Numerical integrations are unstable for \mbox{$\alpha<2.3$}.
}\label{fig:al}
\end{figure}

Fig.~\ref{fig:al} shows that the interference correlation in the MPP depends on the path loss exponent $\alpha$, while it is independent of $\alpha$ for PPPs. This dependence is, however, very small: For low $\alpha$ (close to $2$), the correlation is slightly smaller than for higher values. For values $\alpha\geq 3$, there is almost no change in correlation when further increasing $\alpha$.

\subsection{Impact of {Sensing Range}}

Fig.~\ref{fig:h} shows that the interference correlation decreases with increasing hard-core distance $\range$ and eventually vanishes for $\range\to\infty$. The reason for this behavior is that $\range$ determines the number of senders. A higher $\range$ models a more sensitive sensing, which implies fewer simultaneously sending nodes that are further apart. If a higher fraction of nodes send, naturally the temporal correlation is higher, and vice versa. This is already known for PPPs with a linear relation between fraction of senders and correlation~\cite{ganti09:interf-correl}. For MPPs, we have the same qualitative behavior but with a non-linear relation between fraction of senders and correlation.

\begin{figure}[t]
\centering
\begin{tikzpicture}
\begin{axis}[xlabel={Hard-core distance $\range$},ylabel={Interference correlation $\cor{\interf_1}{\interf_2}$},ymin=0,ymax=1.05,xmin=0,xmax=3,grid=both,xtick={0,0.5,1,1.5,2,2.5,3},
            ytick={0,0.2,0.4,0.6,0.8,1.0},
xticklabels={0,0.5,1,1.5,2,2.5,3},
yticklabels={0,0.2,0.4,0.6,0.8,1.0},
legend style={at={(0.98,0.98)}, anchor=north east, font=\footnotesize},
legend cell align=left
]

\addplot plot[color=black,solid,no marks,style=thick]  table[x index=0,y index=21]  {cor_h.txt};\addlegendentry{MPP $m\rightarrow\infty$}
\addplot plot[color=black,dashed,no marks,style=thick]  table[x index=0,y index=12]  {cor_h.txt};\addlegendentry{MPP $m=6$}
\addplot plot[color=black,dotted,no marks,style=thick]  table[x index=0,y index=4]  {cor_h.txt};\addlegendentry{MPP $m=2$}
\addplot plot[color=black,dashdotted,no marks,style=thick]  table[x index=0,y index=2]  {cor_h.txt};\addlegendentry{MPP $m=1$}
\addplot plot[color=black,loosely dashed,no marks,style=thick]  table[x index=0,y index=1]  {cor_h.txt};\addlegendentry{MPP $m=\frac{1}{2}$}

\end{axis}
\end{tikzpicture}
\caption{Temporal correlation of interference over different values of the hard-core distance $\range$. Parameters are $\densppp=1$ and $\alpha=3$. The glitches in the curves (e.g., close to $\range=2$) are due to instabilities in numerical integrations.}\label{fig:h}
\end{figure}
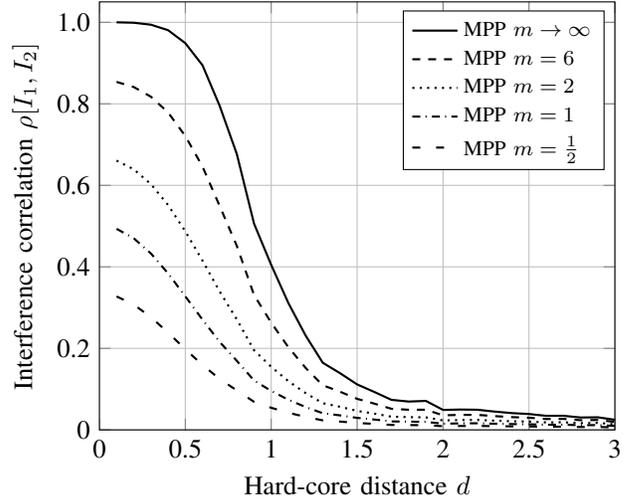
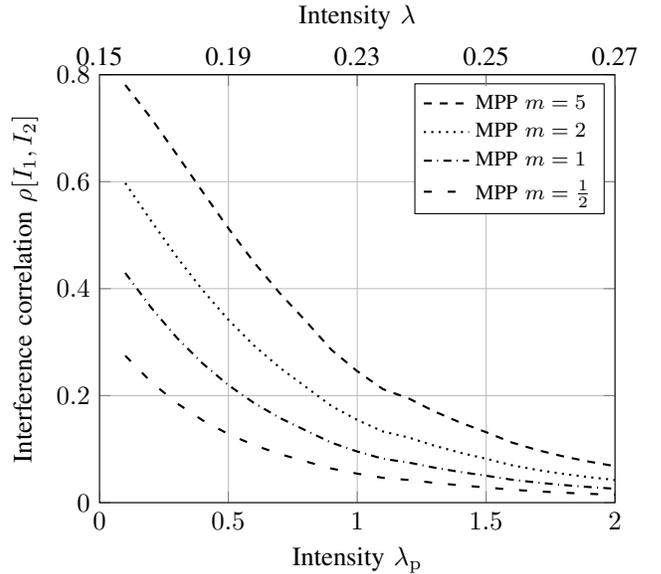
\begin{figure}[t]
\centering
\begin{tikzpicture}
\begin{axis}[xlabel={Intensity $\densppp$},ylabel={Interference correlation $\cor{\interf_1}{\interf_2}$},ymin=0,ymax=0.8,xmin=0,xmax=2,grid=both,
legend style={at={(0.98,0.98)}, anchor=north east, font=\footnotesize},
legend cell align=left
]

\addplot plot[color=black,dashed,no marks,style=thick]  table[x index=0,y index=11]  {cor_lam.txt};\addlegendentry{MPP $m=5$}
\addplot plot[color=black,dotted,no marks,style=thick]  table[x index=0,y index=5]  {cor_lam.txt};\addlegendentry{MPP $m=2$}
\addplot plot[color=black,dashdotted,no marks,style=thick]  table[x index=0,y index=3]  {cor_lam.txt};\addlegendentry{MPP $m=1$}
\addplot plot[color=black,loosely dashed,no marks,style=thick]  table[x index=0,y index=2]  {cor_lam.txt};\addlegendentry{MPP $m=\frac{1}{2}$}

\end{axis}

\begin{axis}[xlabel={Intensity $\dens$},ymin=0,ymax=0.8,xmin=0,xmax=2,grid=none, axis x line*=top, hide y axis,xtick=data,xticklabels from table={cor_lam.txt}{1}]
\end{axis}

\end{tikzpicture}
\caption{Temporal correlation of interference over different values of the intensity $\densppp$ of the PPP to which Mat\'ern thinning is applied. Parameters are $\range=1$ and $\alpha=3$. The upper axis label shows the intensity $\dens$ of the MPP. It monotonically increases with $\densppp$ and is upper bounded by~$\frac{1}{\range^2\pi}$.
}\label{fig:int}
\end{figure}

Overall, we conclude that sensing sensitivity determines the interference correlation: If the sensing is very sensitive, few nodes are sending, thus the correlation is small. If the sensing is nonsensitive, correlation increases up to the point where there is no sensing, which eventually yields slotted ALOHA. In mathematical terms, this is modeled by $\range\to 0$ leading to a PPP.

\subsection{Impact of Intensity}

Fig.~\ref{fig:int} plots the interference correlation of a Mat\'ern network over the intensity $\densppp$ of the PPP from which the MPP is derived from.
It shows a strong decrease of correlation for increasing $\densppp$. This is in contrast to Poisson networks, where the intensity has no impact on interference correlation.
The main reason for this dependency in MPPs is that  $\range$ does not scale with $\densppp$. Hence, for a higher density, a smaller fraction of nodes is allowed to send, since they are on average closer packed, which reduces the retainment probability $\pa$. Indeed, from~\eqref{eq:pa}, we can conclude that $\pa$ monotonically decreases with increasing $\densppp$.

\section{Conclusions}\label{sec:conclusion}

This article contributes to interference calculus in wireless networks with emphasis on the dynamics of interference in Mat\'ern networks with Nakagami fading. We derived and analyzed previously unknown expressions for the variance and covariance of interference power and calculated the correlation coefficient.

We proved that the interference dynamics is significantly different in networks with carrier sensing than in networks without sensing. An important difference is that the interference correlation in Mat\'ern networks depends on the intensity of the underlying point process, which is irrelevant in Poisson networks. The path loss exponent has almost no influence on interference correlation in both types of networks. These results demonstrate the limits of the commonly used Poisson network model. 
At the same time, our results highlight the potential of the Mat\'ern point process as a viable model for networks with carrier sensing: it approximates important aspects of CSMA while remaining tractable to a certain~extend.

\section*{Acknowledgments}
This work has been supported by the Austrian Science Fund (FWF) under grant P24480-N15 (Dynamics of Interference in Wireless Networks). It has also been supported by the K-project DeSSnet, which funded within the context of COMET -- Competence Centers for Excellent Technologies by the Austrian Ministry for Transport, Innovation and Technology (BMVIT), the Federal Ministry for Digital and Economic Affairs (BMDW), and the federal states of Styria and Carinthia. The program is conducted by the Austrian Research Promotion Agency~(FFG).

\begin{IEEEbiography}
    [{\includegraphics[width=1in,height=1.25in,clip,keepaspectratio]{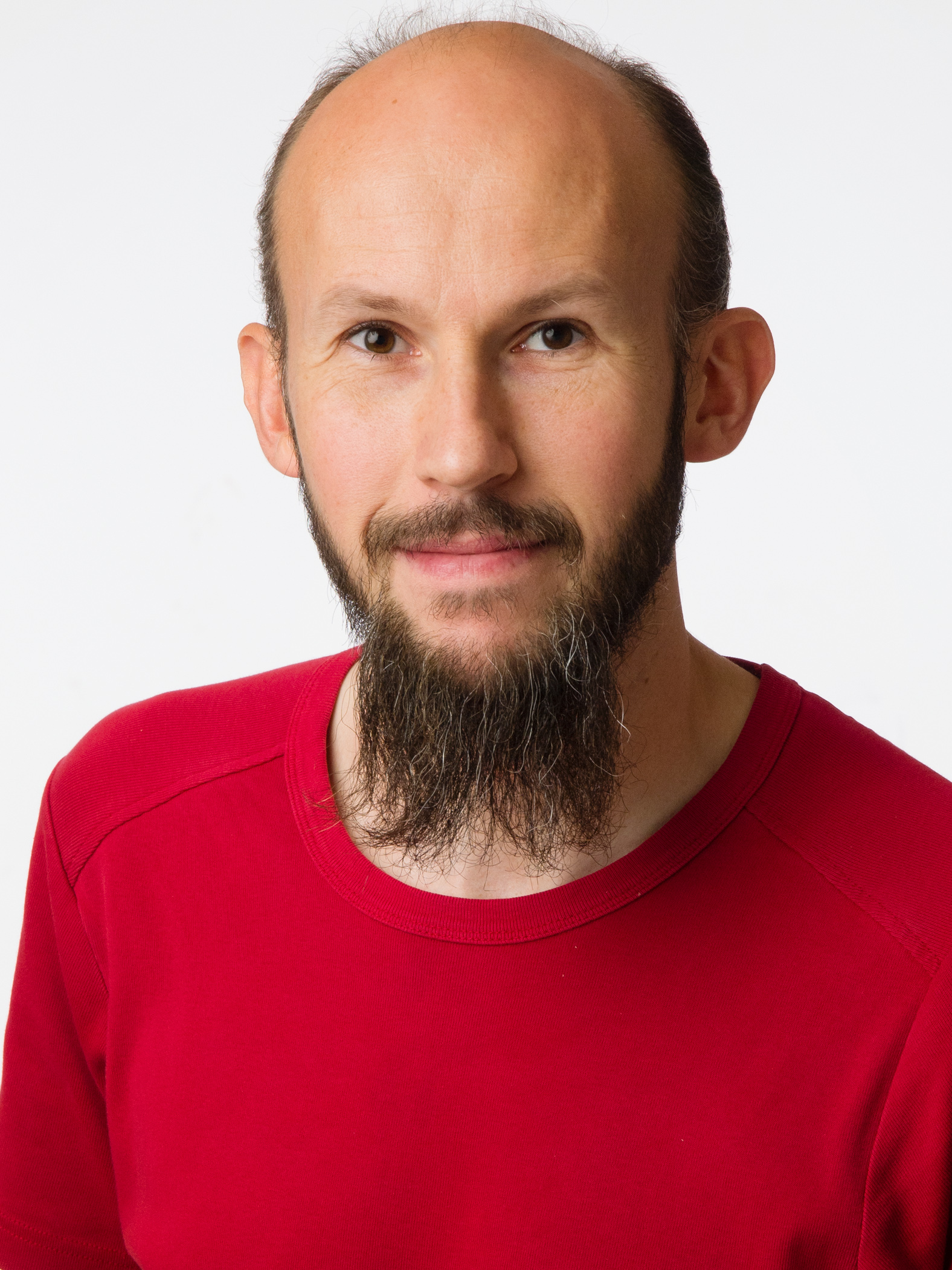}}]{Udo Schilcher}
studied applied computing and
mathematics at the University of Klagenfurt,
where he received two Dipl.-Ing. degrees with
distinction (2005, 2006). From 2005 to 2017, he was
research staff member at the Institute of Networked
and Embedded Systems at the University
of Klagenfurt. His doctoral thesis on inhomogeneous
node distributions and interference correlation in
wireless networks and has been awarded with
a Dr. techn. degree with distinction in 2011. 
After his graduation, from 2011 he was Post-Doc,
again at the Institute of Networked
and Embedded Systems at the University
of Klagenfurt.
Since 2016 he has been senior researcher
at Lakeside Labs GmbH.
His main interests are interference dynamics
and spatial node distributions in wireless networks,
and stochastic geometry.
He received a best paper
award from the IEEE Vehicular Technology Society.
\end{IEEEbiography}

\begin{IEEEbiography}
    [{\includegraphics[width=1in,height=1.25in,clip,keepaspectratio]{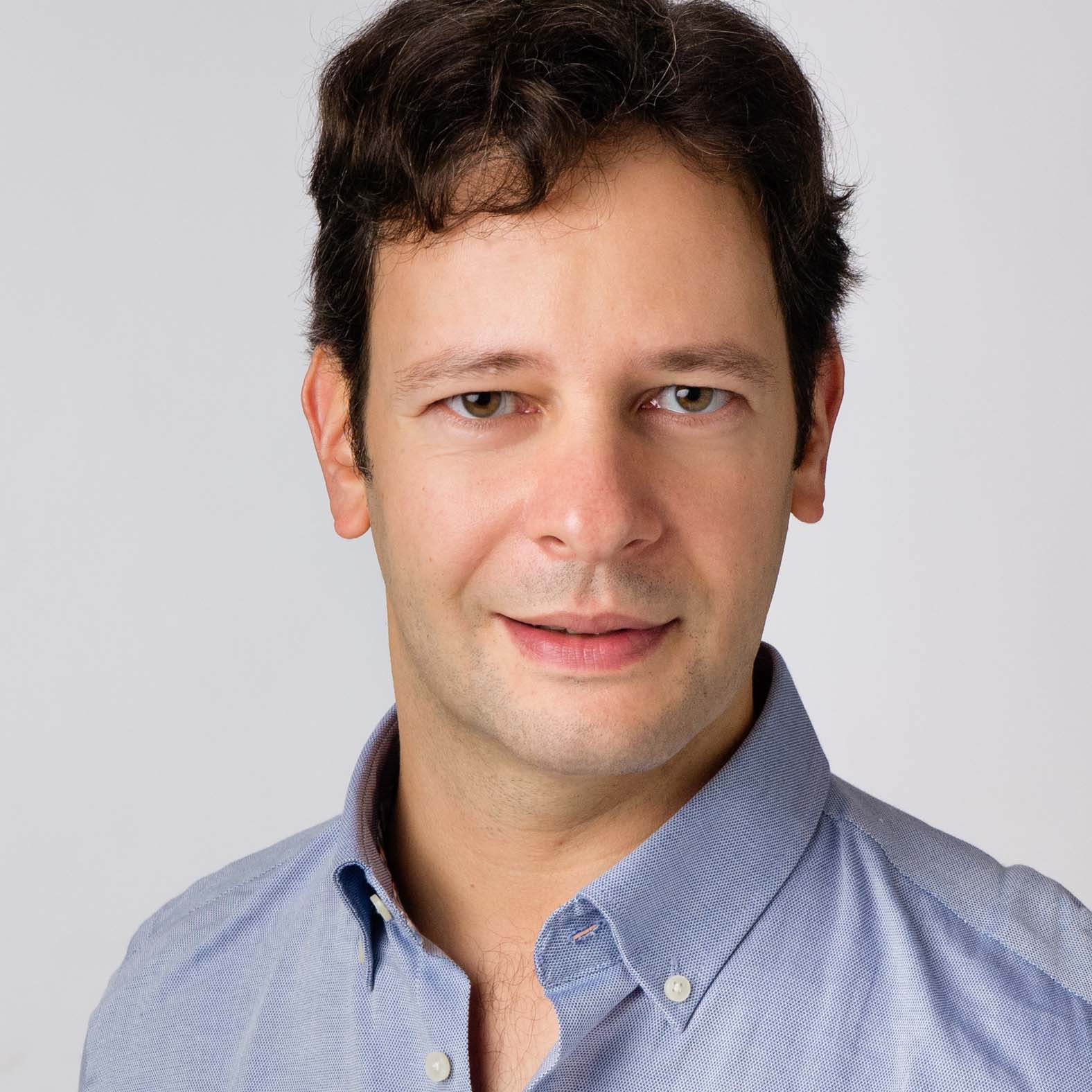}}]{Jorge F. Schmidt}
received the B.Sc. and D.Sc. degrees in electrical engineering from the Universidad Nacional del Sur, Bah\'ia Blanca, Argentina, in 2005 and 2011, respectively. From 2012 to 2014, he was a Postdoctoral Fellow in the Signal Processing and Communications Laboratory at the University of Vigo, Spain. In 2014 he joined the Institute of Networked and Embedded Systems group at University of Klagenfurt, Austria, where he is currently a Senior Researcher. Since 2016 he is also a Senior Researcher at Lakeside Labs GmbH, Austria. His main research interests lie in the area of statistical signal processing and interference modeling and management for wireless communications systems.
He received a best paper award from the ACM SIGSIM.
\end{IEEEbiography}

\begin{IEEEbiography}
    [{\includegraphics[width=1in,height=1.25in,clip,keepaspectratio]{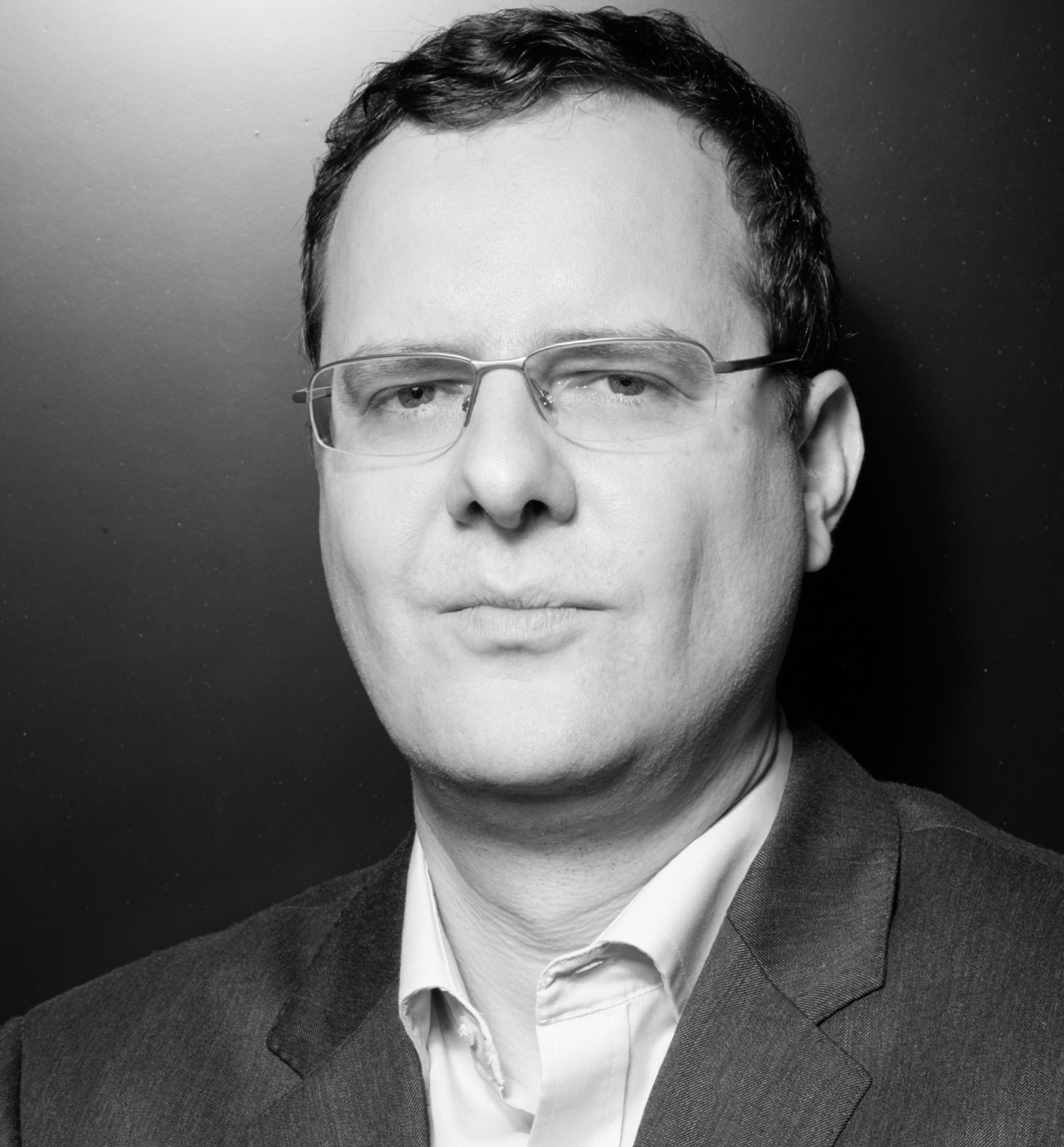}}]{Christian Bettstetter}
(S'98-M'04-SM'09) received the Dipl.-Ing. degree in 1998 and the Dr.-Ing. degree (summa cum laude) in 2004, both in electrical and information engineering from Technische Universit\"at M\"unchen (TUM), Munich, Germany. He was a research and teaching staff member at the Institute of Communication Networks, TUM, until 2003. From 2003 to 2005, he was a senior researcher with DOCOMO Euro-Labs. He has been a professor at the University of Klagenfurt, Austria, since 2005, and founding director of the Institute of Networked and Embedded Systems since 2007. He is also the founding scientific director of Lakeside Labs, a research company on self-organizing networked systems.
\end{IEEEbiography}

\end{document}